\documentclass[reprint,twocolumn,superscriptaddress,secnumarabic,amssymb, nobibnotes, aps, prl]{revtex4-1}
\usepackage[english]{babel}
\usepackage[utf8x]{inputenc}
\usepackage[T1]{fontenc}
\usepackage{comment}

\usepackage{amsmath}
\usepackage{graphicx}
\graphicspath{{Figures/}}
\usepackage{physics}
\usepackage{siunitx}
\usepackage{bm}
\usepackage[colorinlistoftodos]{todonotes}

\newcommand{\bi}{\mathbf{i}}

\newcommand{\bk}{\mathbf{k}}

\newcommand{\bR}{\mathbf{R}}

\newcommand{\td}{\tilde{d}}
\newcommand{\mct}{\mathcal{T}}
\newcommand{\mch}{\mathcal{H}}
\newcommand{\mcv}{\mathcal{V}}
\newcommand{\mcw}{\mathcal{W}}
\newcommand{\mcj}{\mathcal{J}}

\begin{document}
\title{Collective modes in excitonic magnets: dynamical mean-field study}
\author{D. Geffroy}
\affiliation{Department of Condensed Matter Physics, Faculty of Science, Masaryk University, Kotl\'a\v{r}sk\'a 2, 611 37 Brno, Czechia}
\affiliation{Institute for Solid State Physics, TU Wien, 1040 Vienna, Austria}
\author{J. Kaufmann}
\affiliation{Institute for Solid State Physics, TU Wien, 1040 Vienna, Austria}
\author{A. Hariki}
\affiliation{Institute for Solid State Physics, TU Wien, 1040 Vienna, Austria}
\author{P. Gunacker}
\affiliation{Institute for Solid State Physics, TU Wien, 1040 Vienna, Austria}
\author{A. Hausoel}
\affiliation{Institute for Theoretical Physics and Astrophysics, University of Würzburg, Am Hubland 97074 Würzburg, Germany}
\author{J. Kune\v{s}}
\affiliation{Institute for Solid State Physics, TU Wien, 1040 Vienna, Austria}
\affiliation{Institute of Physics,
Czech Academy of Sciences, Na Slovance 2,
182 21 Praha 8, Czechia}

\begin{abstract}
We present a dynamical mean-field study of dynamical susceptibilities in 
two-band Hubbard model. Varying the model parameters we analyze the two-particle
excitations in the normal as well as in the ordered phase, an excitonic condensate.
The two-particle DMFT spectra in the ordered phase reveal the gapless Goldstone modes 
arising from spontaneous breaking of continuous symmetries. We also observe
gapped Higgs mode, characterized by vanishing of the gap at the phase boundary. 
Qualitative changes observed in the spin susceptibility can be used as an experimental 
probe to identify the excitonic condensation.  
\end{abstract}

\maketitle
Long-rang order (LRO) and the concomitant spontaneous symmetry breaking are a prominent demonstration of collective behavior in solids. For common LROs, e.g., magnetic order, the order parameter can be observed directly with present technology. However, some exotic LROs, dubbed hidden orders, have been recognized through their thermodynamic properties so far~\cite{Kung2015,tsubouchi2002,Lu2017}. In this case, dynamical response functions are invaluable for understanding the nature of the LRO. Excitonic insulator~\cite{Mott1961,Halperin1968b} is an example of LRO that after decades of defying detection has been identified through its dynamical fingerprint~\cite{Kogar2017}. Recently, realization of excitonic magnet~\cite{Khaliullin2013}, an analog of excitonic insulator arising by condensation of spinful excitons, was proposed in Pr$_{0.5}$Ca$_{0.5}$CoO$_3$~\cite{tsubouchi2002,Kunes2014b,Yamaguchi2017} with conclusive evidence still missing. Excitonic interpretation of magnetism of Ca$_2$RuO$_4$ is subject of current debate~\cite{Jain2017,Zhang2017}. Excitonic condensation finds its formal analogy in the
LRO in Heisenberg-dimer systems, experimentally studied in TlCuCl$_3$~\cite{Merchant2014}.

In this Letter we use dynamical mean-field theory (DMFT) to study the dynamical response of an excitonic magnet realized in two-band Hubbard model.
We show that the symmetry lowering gives rise to a coupling between the
exciton condensate fluctuations and spin, which leaves a unique signature in the
dynamical spin structure factor and thus can be observed by experiments such as inelastic neutron or x-ray scattering.

Breaking of continuous symmetry in systems with short range interactions
results in appearance of gapless Goldstone modes (GMs)~\cite{Watanabe2012} as well as
gapped Higgs excitations~\cite{Pekker2015}, which have been the subject of recent interest
~\cite{Jain2017,Sherman2015,Merchant2014,Pekker2015}.
The description of interacting electron systems typically relies on
perturbative approaches in the weak-coupling
limit~\cite{Bascones2002,Yamaguchi2017} or low-energy effective models
in the strong-coupling limit~\cite{Khaliullin2013,Tatsuno2016}.
The dynamical mean-field theory (DMFT)~\cite{Metzner1989,Georges1996}
uses a different approximation assuming a Luttinger-Ward functional
with the local propagators only, which
makes DMFT tractable for realistic multi-orbital models~\cite{Kotliar2006,Held2007}. DMFT 
has been widely used to study one-particle (1P) dynamics, while
two-particle (2P) susceptibilities are rarely studied for multi-orbital
models~\cite{Kunes2011,Park2011,Boehnke2013,Kunes2014a,Hoshino2016,DelRe2017}
and their behavior in ordered phases is unexplored.
\begin{figure}
  \begin{tabular}{cc}
    \includegraphics[width=0.41\columnwidth]{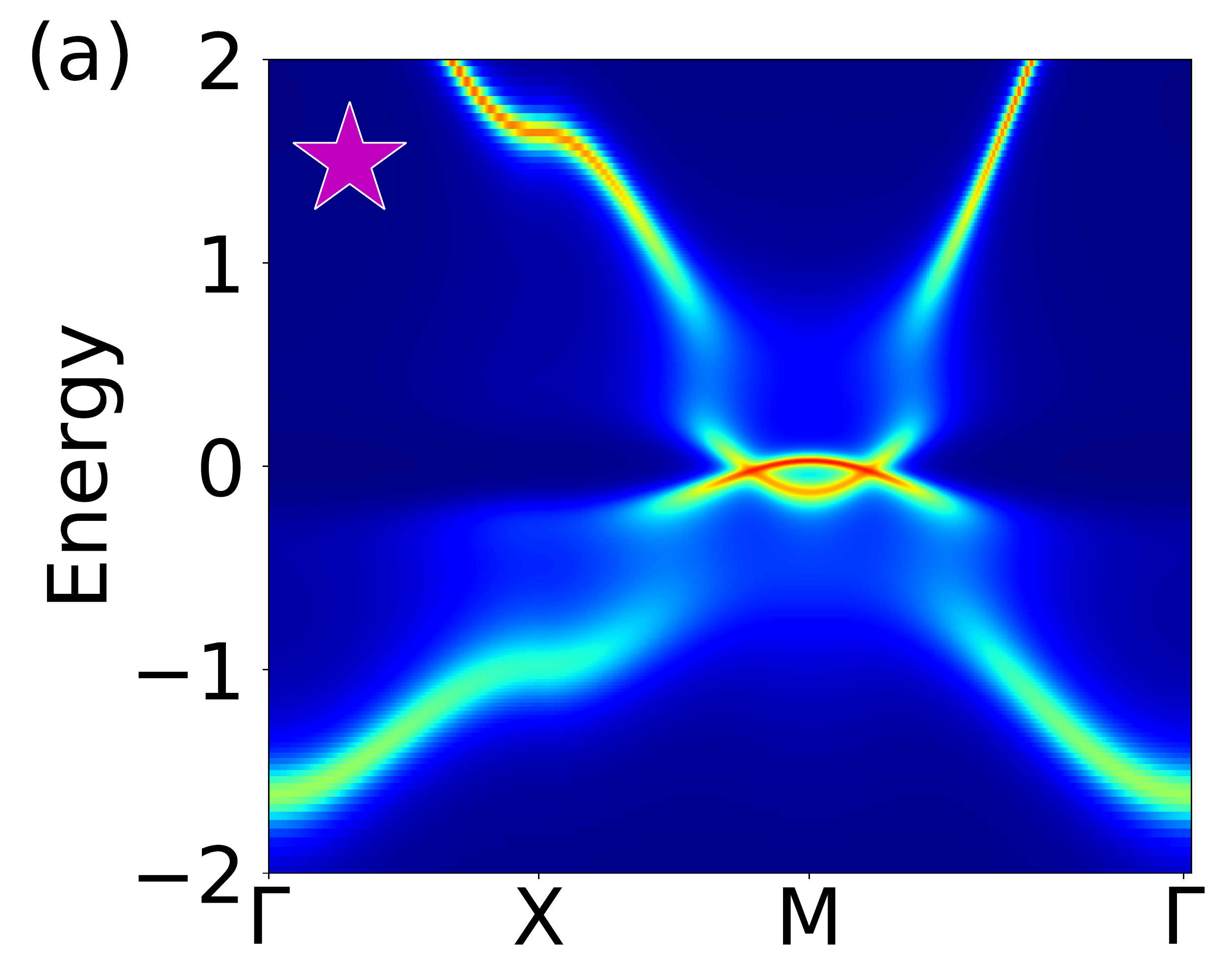} &
    \includegraphics[width=0.4\columnwidth]{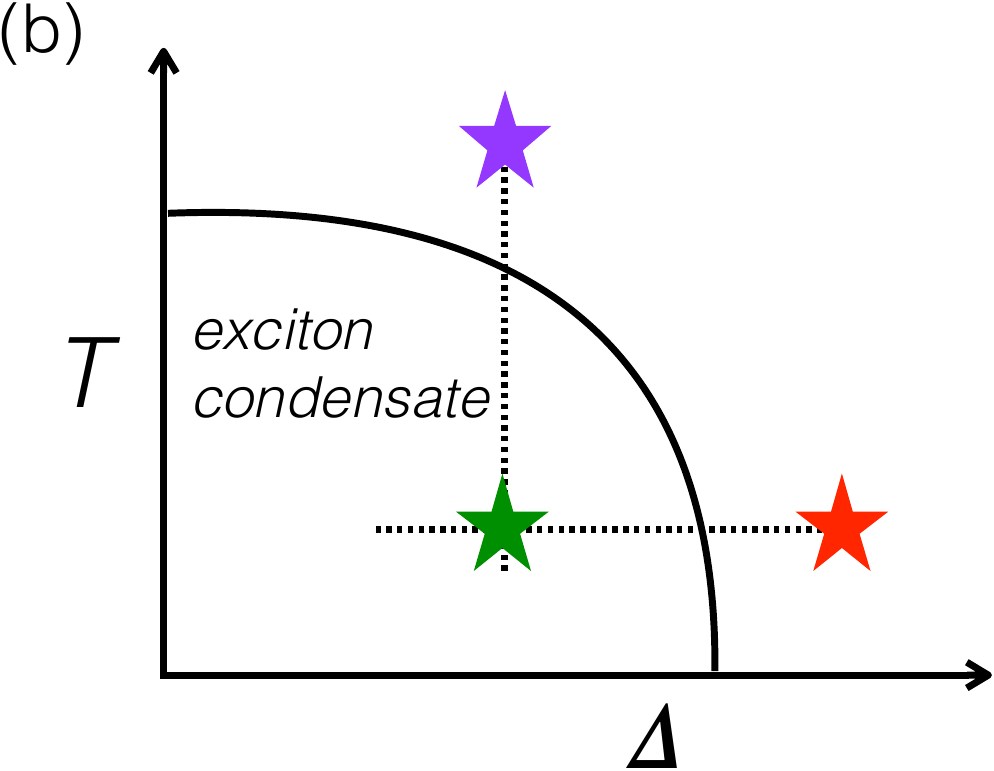} \\
    \includegraphics[width=0.41\columnwidth]{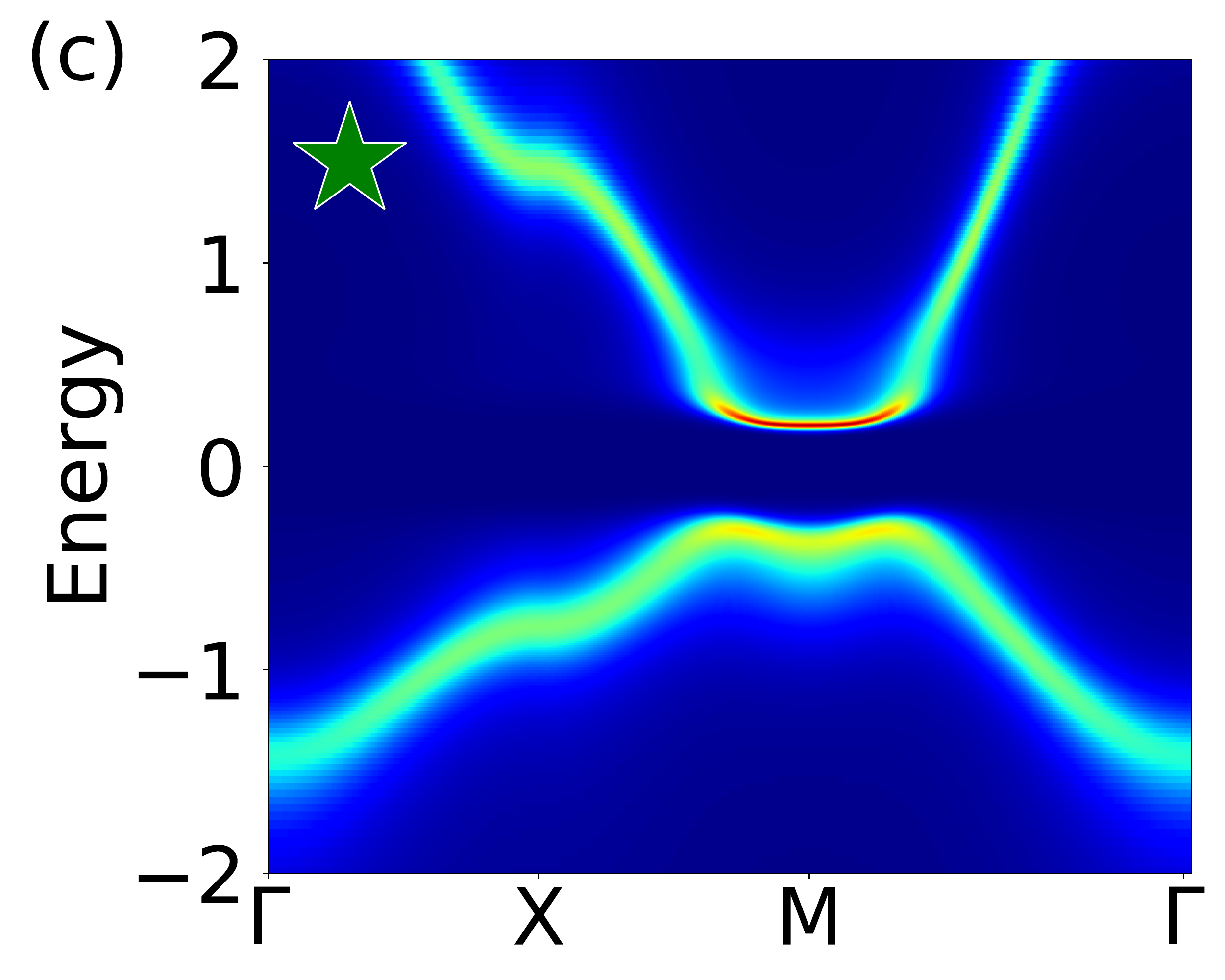} &
    \includegraphics[width=0.49\columnwidth]{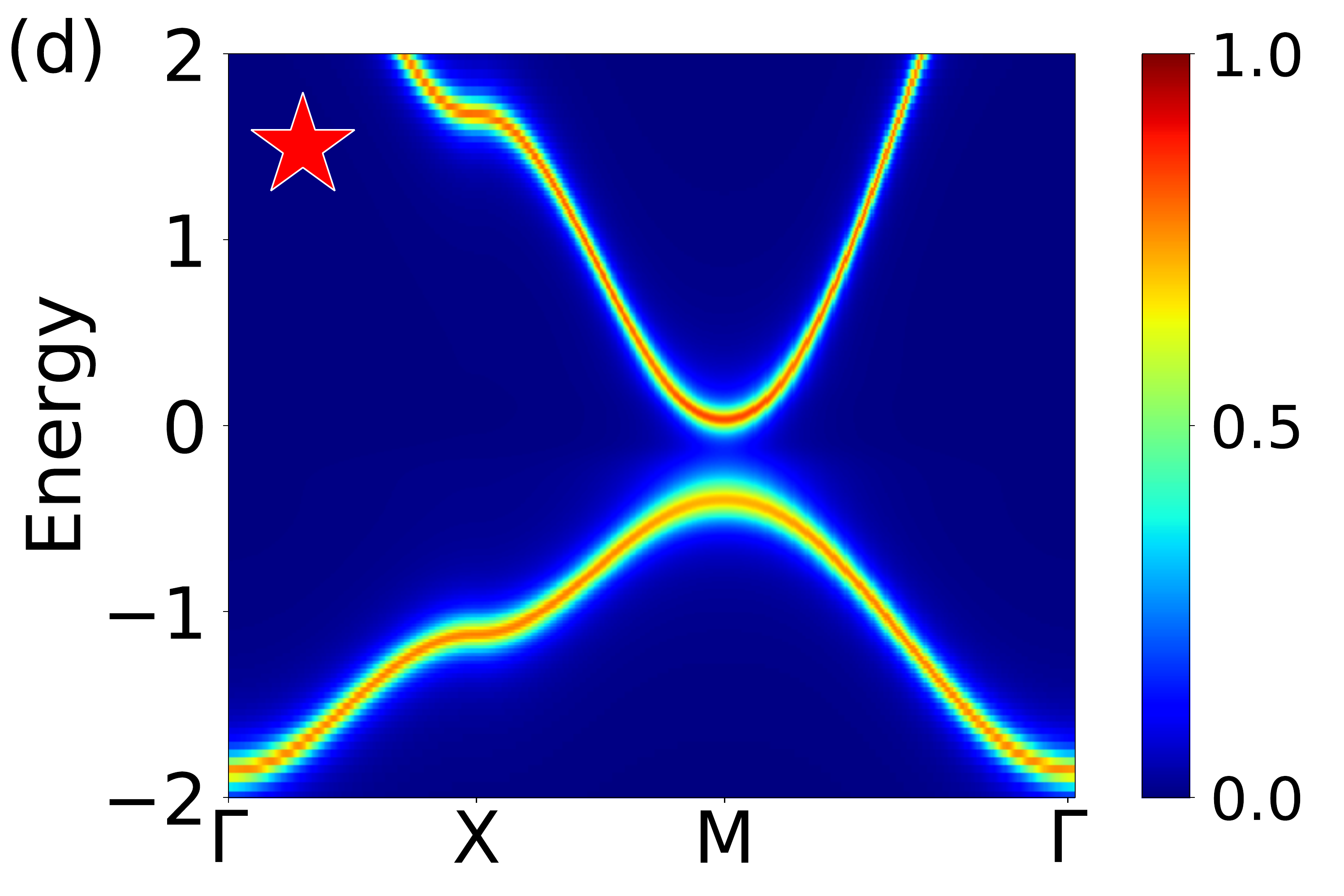} 
  \end{tabular}
  \caption{\label{fig:1p_disp} Sketch of the cryatal field vs temperature ($\Delta-T$) phase diagram (b) with
  marked cuts, along which the susceptibilities are calculated.
  The 1P spectral function at the $(\Delta,T)$-points violet (3.55, 1/11),
  green (3.55, 1/40) and blue (3.8, 1/40).}
\end{figure}


The Hamiltonian of studied model reads
\begin{align}
    H=&\sum_{ij,\sigma} 
    \begin{pmatrix} a_{i\sigma}^{\dagger}  & b_{i\sigma}^{\dagger} \end{pmatrix}
    \!
    \begin{pmatrix} t_{aa} & 
    t_{ab}\\ t_{ab} &\ t_{bb} \end{pmatrix}
    \!
    \begin{pmatrix} a_{j\sigma}^{\phantom\dagger}  \\ b_{j\sigma}^{\phantom\dagger} \end{pmatrix}
    +\frac{\Delta}{2}\sum_{i,\sigma}(n^a_{i\sigma}-n^b_{i\sigma})
    \nonumber
    \\
    &+ U \sum_{i,\alpha}n^\alpha_{i\uparrow}n^\alpha_{i\downarrow}+
    \sum_{i,\sigma\sigma'}(U'-J\delta_{\sigma\sigma'}) n^a_{i\sigma}n^b_{i\sigma'}, 
    \label{eq:2bhm}
\end{align}
where $a^{\dag}_{i\sigma}$ and $b^{\dag}_{i\sigma}$ are fermionic
operators that create electrons with the respective orbital flavors
and spin $\sigma$ at site $i$ of a square lattice. The first term
describes the nearest neighbor hopping.
The rest, expressed in terms of local densities
$n^{c}_{i,\sigma} \equiv c^{\dag}_{i\sigma}c_{i\sigma}$,
captures the crystal-field $\Delta$, the Hubbard
interaction $U$ and Hund's exchange $J$ in the Ising approximation. Parameters
$U=4$, $J=1$, $U'=U-2J$,~\footnote{The results are little sensitive to variation of $U'$ and $J$ as long
as the ratio $\Delta/J$ is fixed.} $t_{aa}=0.4118$, $t_{bb}=-0.1882$, $t_{ab}=$0, 0.02, 0.06 
with magnitudes (in eV) typical for $3d$ transition metal oxides 
were used in previous studies~\cite{Kunes2014a,Kunes2014c,Kunes2016}. 

We follow the standard DMFT procedure of self-consistent mapping
the lattice model onto an auxiliary Anderson impurity model (AIM)~\cite{Georges1992,Jarrell1992},
which is solved with the ALPS implementation~\cite{Bauer2011, Shinaoka2016, Gaenko2017}~\footnote{Numerically identical results for the normal state susceptibilities were obtained
with w2dymanics~\cite{w2dynamics}.}
of the matrix version of the strong-coupling continuous-time quantum
Monte-Carlo (CT-QMC) algorithm~\cite{Werner2006a}.
The susceptibilities~\cite{Georges1996,Kunes2011,vanLoon2015,Krien2017} are obtained by solving 
the Bethe-Salpeter equation in the particle-hole channel with the DMFT 1P propagators 
and 2P-irreducible vertices of AIM using the orthogonal polynomial representation~\cite{Boehnke2011}.

The susceptibilities $\chi^{OO}_{\eta\eta}(\bk,\omega)$ are obtained by analytic
continuation~\cite{Gubernatis1991,SM} of their Matsubara representations
\begin{equation*}
\chi^{OO}_{\eta\eta}(\bk,i\nu_n)=
\!
\sum_{\bR}
\!
\int_0^{\beta}\!\!\!\!\!\mathrm{d}\tau 
\,e^{i(\nu_n\tau+\bk\cdot\bR)} 
\langle O^{\eta}_{\bi+\bR}(\tau)O^{\eta}_\bi(0)
\rangle
-\langle O^\eta\rangle^2\!\!,
\end{equation*}
 with the observables $O$ of interest being excitonic fields
 $R^\eta_i(I^\eta_i)=\sqrt{\pm 1}\sum_{\alpha\beta}
\sigma^\eta_{\alpha\beta}(
a_{i\alpha}^{\dagger} b_{i\beta}^{\phantom\dagger}\pm
b_{i\alpha}^{\dagger} a_{i\beta}^{\phantom\dagger})$, respectively,  
with $\eta=x,y$ and the $z$-component of spin moment
${S^z_i=\sum_{\alpha\beta}\sigma^z_{\alpha\beta}
(a_{i\alpha}^{\dagger} a_{i\beta}^{\phantom\dagger}
+
b_{i\alpha}^{\dagger} b_{i\beta}^{\phantom\dagger})}$.

\begin{figure}[t]
\includegraphics[width=0.48\textwidth]{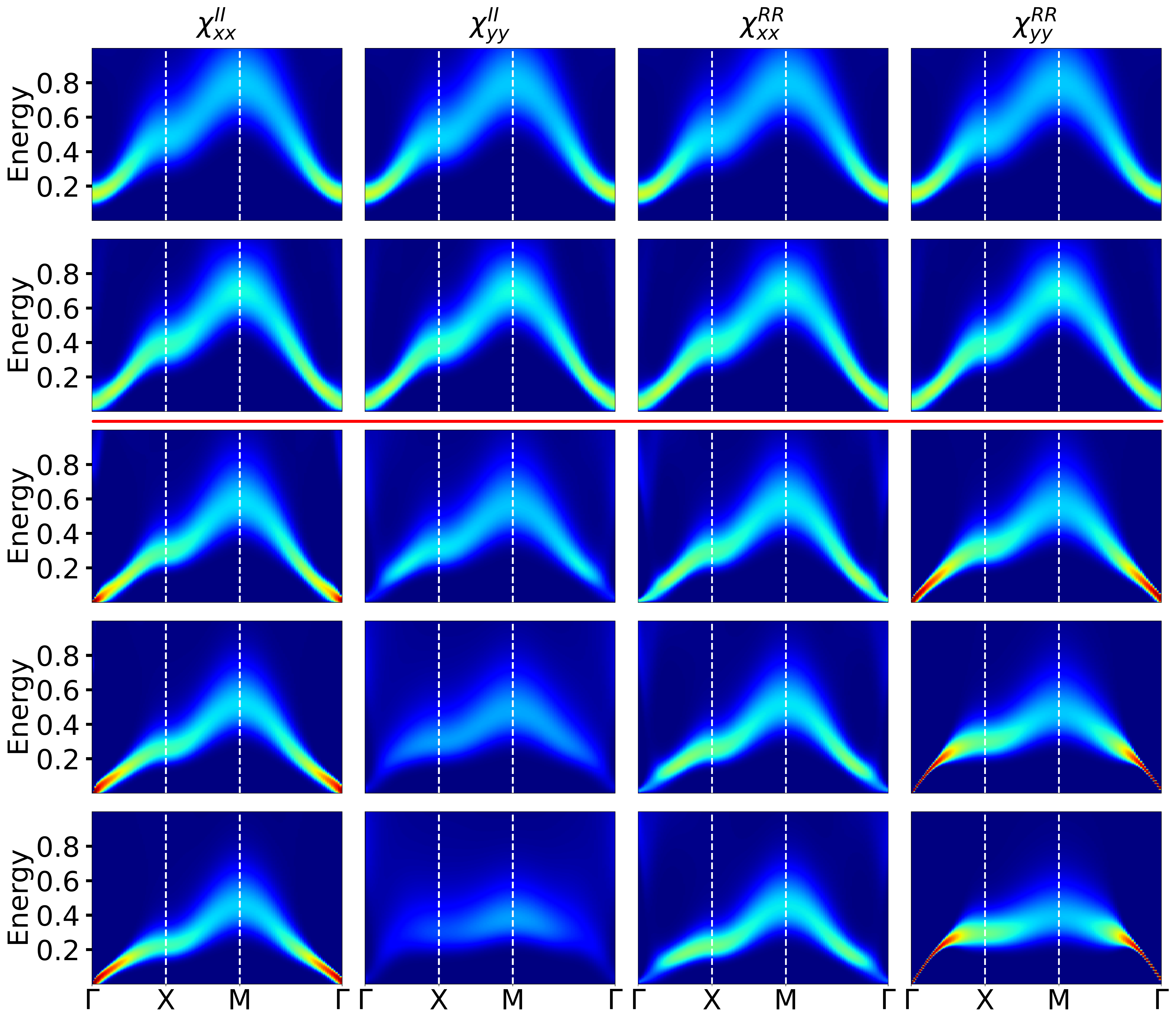}
\caption{\label{fig:goldstone}Evolution of the excitonic modes of 
  dynamical susceptibility in the $U^2(1)$ model
  ($t_{ab}$=0) across $\Delta$-driven transition ($T=1/40$). The columns correspond to 
  $-\operatorname{Im}\chi^{OO}_{\gamma\gamma}(\bk,\omega)$ with $O^\gamma=I^x,I^y,R^x,R^y$ (left to
  right) along the high-symmetry lines in the 2D Brillouin zone. The
  rows from top to bottom correspond to $\Delta=$3.9, 3.8, 3.65, 3.55,
  3.45 with $\Delta_c\approx 3.75$ (Red line separates the normal state
  from the PEC phase).}
\end{figure}
\begin{figure}[t]
\includegraphics[width=0.45\columnwidth]{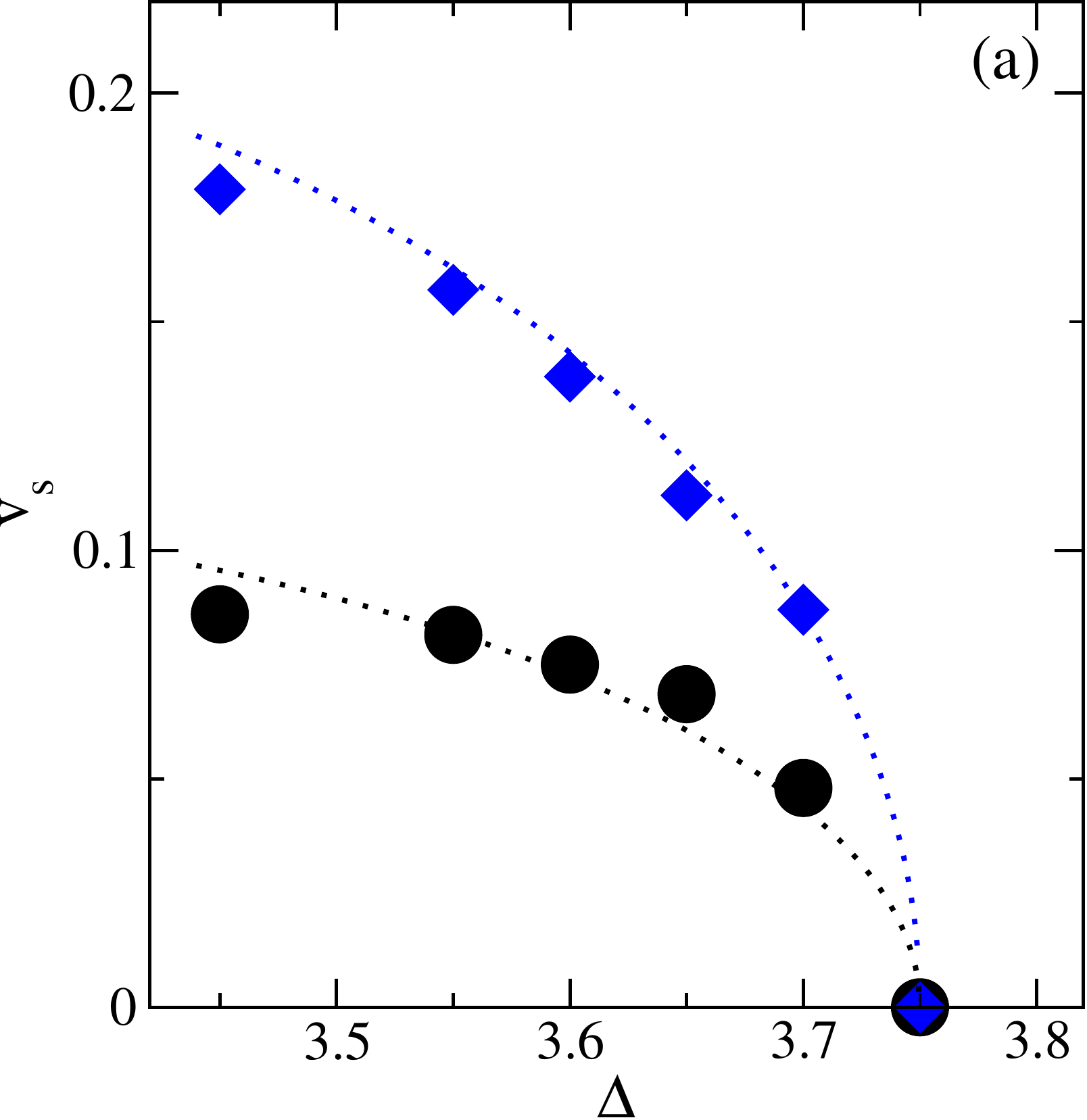}
\includegraphics[width=0.45\columnwidth]{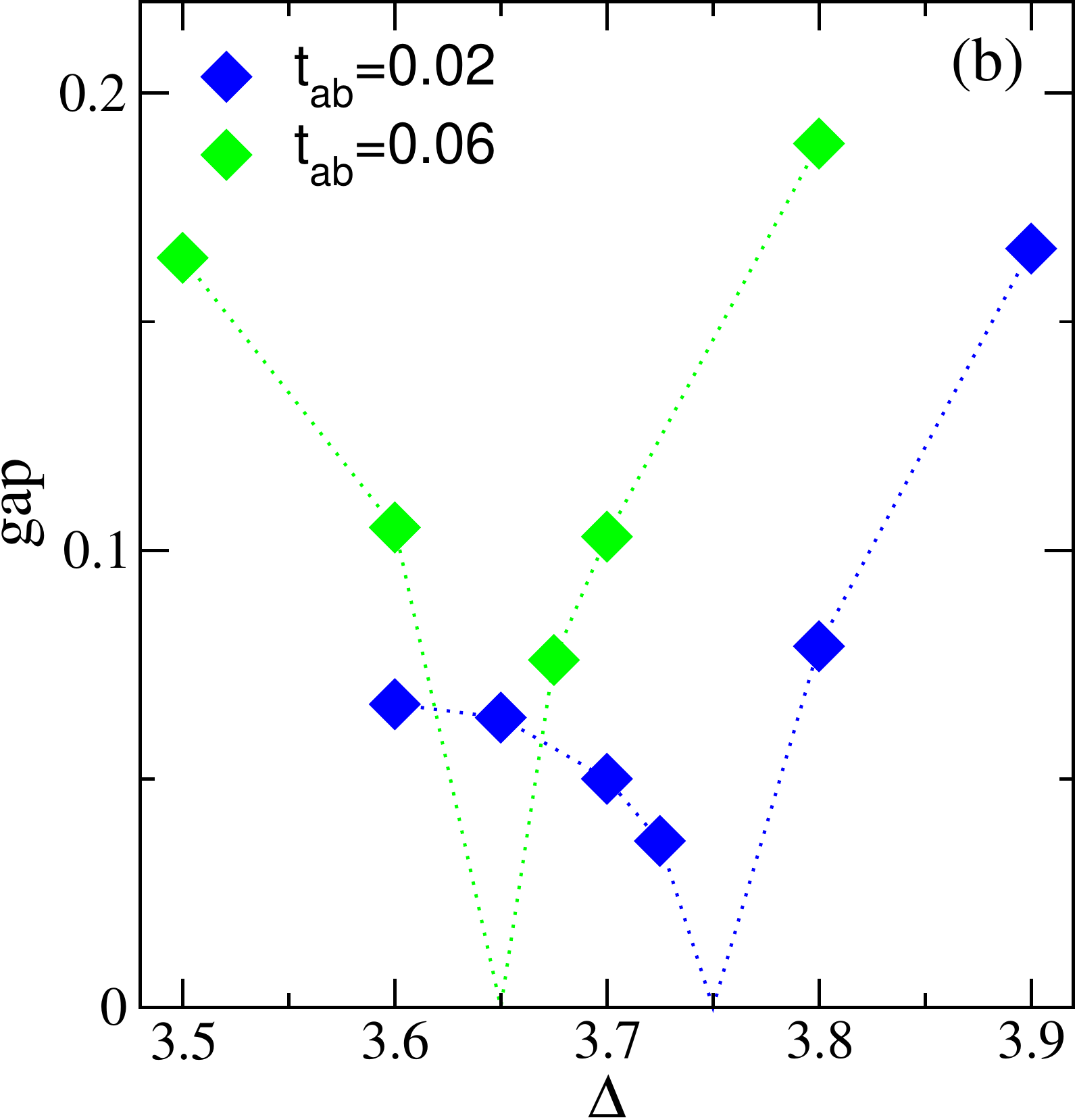}
\caption{\label{fig:gap} (a) The sound velocity $v_s$ of the GMs, the phase mode  ($\chi^{RR}_{yy}$, blue symbols) and the spin rotation mode ($\chi^{II}_{xx}$, black symbols)
in the $U^2(1)$ model as a function of the crystal field $\Delta$. The dotted lines show the
corresponding strong-coupling results.
(b) The Higgs gap in the $U(1)$ model with $t_{ab}=$0.02 and 0.06 as a function of $\Delta$. The line
is a guide for eyes.}
\end{figure}

Model (\ref{eq:2bhm}) at half-filling has a rich phase diagram
exhibiting a metal-insulator transition~\cite{Werner2007} as well as
various types of LRO including antiferromagnetism,
spin-state order or
superconductivity~\cite{Kaneko2014,Kaneko2015,Kunes2014a,Hoshino2016}. For
the present parameters it undergoes a temperature- or crystal-field-controlled transition to polar exciton condensate (PEC)~\cite{Kunes2015}, as shown in Fig.~\ref{fig:1p_disp}b.
PEC is characterized by a finite excitonic field. Throughout the paper we choose the orientation $\langle I^y\rangle=\phi$, while $R^y$, $I^x$ and $R^x$ remain fluctuating. This phase is an instance of spin nematic state, which breaks spin-rotation symmetry without appearance of spin polarization.

The behavior of the collective modes depends on the continuous symmetry broken by the LRO~\cite{Watanabe2012}. Here, it is the $U(1)$ spin ($z$-axis) rotation. If $t_{ab}=0$, an additional $U(1)$ gauge symmetry due to conservation of $\sum_{i,\sigma}(n^a_{i,\sigma}-n^b_{i,\sigma})$ makes the total broken symmetry $U(1)\times U(1)$. We will refer to the general $t_{ab}\neq 0$ case as $U(1)$ model and the $t_{ab}=0$ case as $U^2(1)$ model.

{\it $\Delta$-driven transition.} While the system exhibits a sizable 1P gap throughout the studied $\Delta$-range, horizontal line in Fig.~\ref{fig:1p_disp}b, low-energy 2P-excitations show up
in the excitonic susceptibilities, Fig.~\ref{fig:goldstone}. In the normal phase ($\Delta>\Delta_c$), these can be viewed as spinful Frenkel excitons. The spin symmetry ensures the equivalence of $x$ and $y$
directions, while the gauge symmetry leads to equivalence of 
the excitonic fields $R$ and $I$ in the $U^2(1)$ model.

Reducing $\Delta$ closes the excitation gap and the system undergoes transition to the PEC phase.
For the exctionic field, which freezes in an arbitrary direction both in the $xy$-plane and the $RI$-plane in the $U^2(1)$ case, we choose the orientation discussed above.
Linear gapless GMs~\cite{SM} corresponding to the spin rotation and phase fluctuation 
($RI$-rotation) are observed in $\chi^{II}_{xx}$ and $\chi^{RR}_{yy}$, respectively.
The intensities of both GMs diverge as $1/|\bk|$~\cite{SM}.
The corresponding sound velocities are shown in Fig.~\ref{fig:gap}a.
\begin{figure}[t]
  \centering
  \vspace{-0.4cm}
  \includegraphics[width=0.99\columnwidth]{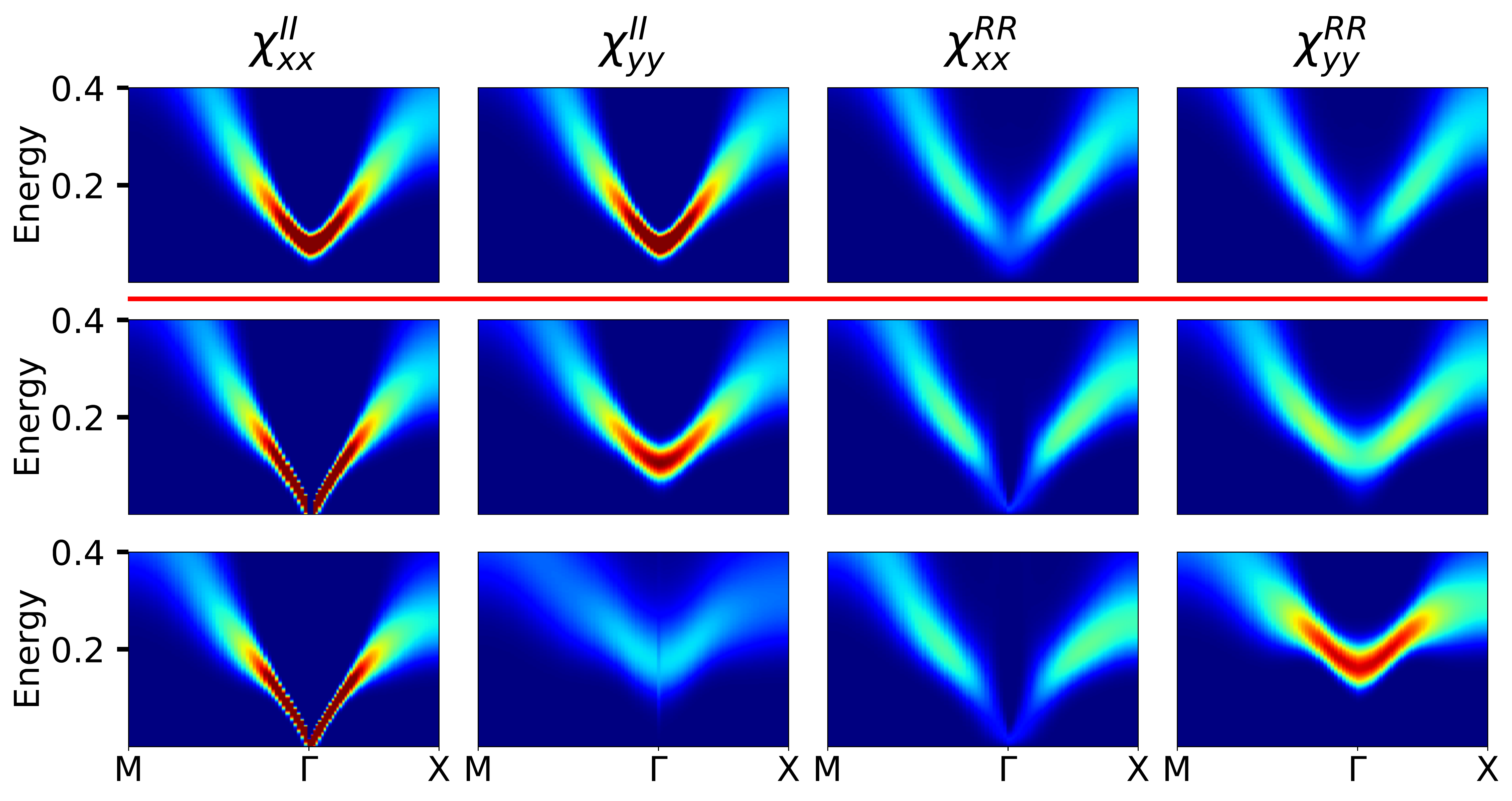}
\caption{\label{fig:higgs} 
The same susceptibilities as in Fig.~\ref{fig:goldstone} ($T=1/40$)
in the vicinity of $\Gamma$-point for $U(1)$ model with cross-hopping
$t_{ab}=0.06$. The rows from top to bottom correspond to
$\Delta=$3.675, 3.60, 3.5, with $\Delta_c\approx 3.65$ (Red line separates the normal state
  from the PEC phase) .}
\end{figure}
\begin{figure}[b]
\includegraphics[width=0.5\textwidth]{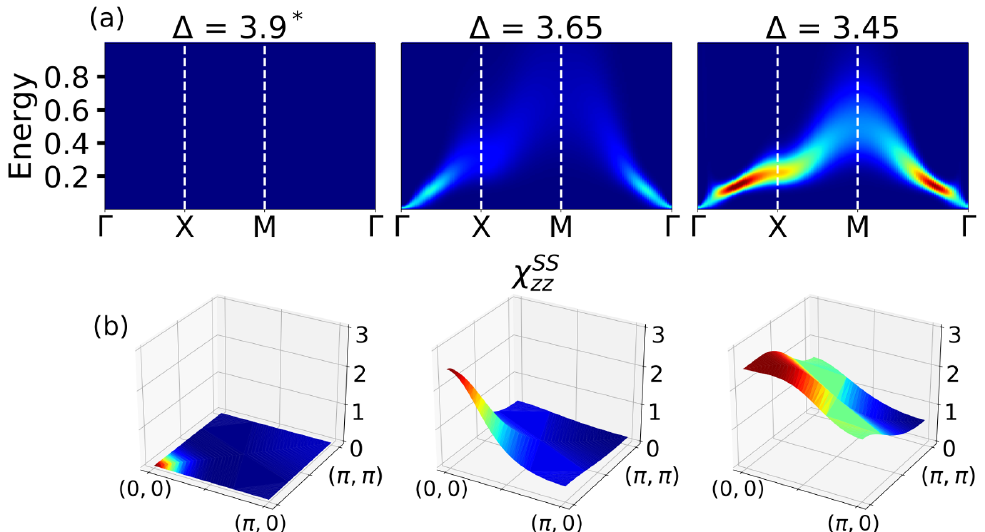}
\caption{\label{fig:spin} (a) Evolution of dynamical spin susceptibility
$-\operatorname{Im}\chi^{SS}_{zz}(\bk,\omega)$ across the $\Delta$-driven transition in the $U^2(1)$ model of Fig.~\ref{fig:goldstone} (Asterisk marks the normal phase). (b) The corresponding static susceptibilities $\operatorname{Re}\chi^{SS}_{zz}(\bk,0)$
throughout the Brillouin zone.}
\end{figure}

Finite cross-hopping $t_{ab}$ leads to a generic $U(1)$ model. The equivalence between the $R$ and $I$
fields is lost, see Fig.~\ref{fig:higgs}. The excitonic field freezes
in the $I$-direction~\cite{Kunes2015,Geffroy2018}, while the $xy$-orientation remains arbitrary.
For the small $t_{ab}$ studied here, the changes to the excitonic spectra~\cite{SM} are concentrated in the
low-energy region shown in Fig.~\ref{fig:higgs}. The spin-rotation GM, visible in 
$\chi^{II}_{xx}$, remains gapless and linear. The 'phase' mode acquires a Higgs gap that vanishes at
the transition, Fig.~\ref{fig:gap}b, a behavior observed in bi-layer Heisenberg system TlCuCl$_3$~\cite{Merchant2014}.

Interestingly the character of this mode changes
as we proceed deeper into the ordered phase, Fig.~\ref{fig:higgs}.
Close to the phase boundary, its spectral weight is dominated by $\chi^{II}_{yy}$, i.e.,
amplitude fluctuation of the condensed $I^y$ field. Deeper in the ordered phase 
the spectral weight is mostly in $\chi^{RR}_{yy}$, corresponding to phase fluctuation ($RI$-rotation)
as in the $U^2(1)$ model. We offer an interpretation in terms of the relative strength of the
symmetry breaking term ($t_{ab}$) in the Hamiltonian and the spontaneously generated Weiss field. The Weiss field, the off-diagonal $F_{ab}^{\uparrow\downarrow}(\omega)$ part of the hybridization function in the present method, 
is in general a fluctuating (frequency dependent) object, which prohibits a direct comparison 
to $t_{ab}$. Nevertheless, we can compare their dynamical effects. A Weiss field dominating over the
Hamiltonian term ($t_{ab}$) results in a gapped GM found deep in the ordered phase. A common example of such situation is a gap in spin-wave spectra of magnets due to magneto-crystalline anisotropy.
Dominance of the Hamiltonian term ($t_{ab}$) close to the phase boundary, where the Weiss field is small,
results in amplitude fluctuations. This is a generic situation in cases without an approximate symmetry.
This interpretation is supported by the observation that the extent of the amplitude-fluctuation
regime shrinks when $t_{ab}$ is reduced~\cite{SM}. Moreover, the strong-coupling calculations (see SM~\cite{SM}), which make an explicit comparison possible, lead to the same conclusions.

Next, we discuss the impact of exciton condensation on the spin susceptibility $\chi^{SS}_{zz}$, shown in Fig.~\ref{fig:spin}. In the normal phase, $\chi^{SS}_{zz}(\bk,\omega)$ exhibits no distinct dispersion and 
essentially vanishes throughout the Brillouin zone, Fig.~\ref{fig:spin}b, as expected
in a band insulator. In the PEC phase, it develops a sharp spin-wave-like dispersion although 
there are no ordered moments present. We point out a similarity of $\chi^{SS}_{zz}(\bk,\omega)$ to $\chi^{RR}_{xx}(\bk,\omega)$ that we discuss later. A distinct feature of $\chi^{SS}_{zz}(\bk,\omega)$
is the suppression of the spectral weight close to the $\Gamma$-point. This suppression can be
overcome by doping, which results in appearance of ferromagnetic exciton condensate~\cite{Kunes2014c}.
\begin{table}[t]
\caption{\label{tab:sw} The parameters of Eq.~\ref{eq:sw}. 
The variational parameter $0\leq\alpha^2\leq 1$, corresponding to the LS density,
assumes 1 in the normal phase and $\tfrac{\mu+z(\mct+\mcw)+z\mcv}{2z(\mct+\mcw)+z\mcv}$ in the condensate.}
\centering
\begin{tabular}{c|l}
 $\mu_x$ & $\alpha^2\mu+z\alpha^2(1-\alpha^2)(2\mct+2\mcw+\mcv)$ \\
  $\mct_x$ & $\alpha^2\mct-(1-\alpha^2)\mcj$ \\
  $\mcw_x$ & $\alpha^2\mcw+(1-\alpha^2)\mcj$ \\ 
  \hline
 $\mu_y$ & $z(\mct+\mcw)$; $\mu$ if $\alpha^2=1$ \\
 $\mct_y$ & $ \mct-\alpha^2(1-\alpha^2)(2\mct+2\mcw+\mcv)$ \\
 $\mcw_y$ & $\mcw-\alpha^2(1-\alpha^2)(2\mct-2\mcw+\mcv)$
 \end{tabular}
\end{table}

{\it Strong-coupling limit.} To understand the numerical results, it is instructive to analyze the
strong-coupling limit of (\ref{eq:2bhm}), which 
can be expressed in terms of two-flavor hard-core bosons~\cite{Balents2000b,Kunes2015,Nasu2016} 
\begin{equation}
\label{CartHam}
\begin{split}
\mch=&\mu\sum_{i} 
n_i
-\!
\sum_{ij,\nu}
\Big[
\mct
d_{i\nu}^{\dagger}d_{j\nu}^{\phantom\dagger}-
\frac{\mcw}{2}
(
d_{i\nu}^{\dagger}d_{j\nu}^{\dagger}+ 
d_{i\nu}^{\phantom\dagger}d_{j\nu}^{\phantom\dagger}
)
\Big]
\\
+&\frac{\mcv}{2}\sum_{ij}
n_in_j
+\frac{\mcj}{2}\sum_{ij}
S^z_iS^z_j,
\end{split}
\end{equation}
Bosonic operators $d_{i\nu}^\dagger$ ($\nu=x,y$), which create high-spin (HS) states out
of the low-spin (LS) state, are related to the excitonic fields by
$R_i^\nu(I_i^\nu)\rightarrow \sqrt{\pm 1}(d_{i\nu}^\dagger\pm d_{i\nu}^{\phantom\dagger})$.
The number operators $n_i=\sum_{\nu}d_{i\nu}^{\dagger}d_{i\nu}^{\phantom\dagger}$ measure
the HS concentration and ${S^z_i=-i(d_{ix}^{\dagger}
  d_{iy}^{\phantom\dagger}-d_{iy}^{\dagger}d_{ix}^{\phantom\dagger})}$
is the $z$-component of the spin operator.
The relations of the coupling constants $\mu$, $\mct$, $\mcw$, $\mcv$, and $\mcj$ to the parameters of (\ref{eq:2bhm}) can be
found in SM~\cite{SM} and Ref.~\onlinecite{Kunes2014a}.
Since $\mcw\sim t_{ab}^2$, the gauge symmetry of the
$U^2(1)$ model reflects conservation of $d$-charge for $\mcw=0$.

Generalized spin wave treatment~\cite{Sommer2001,Nasu2016}
of the excitations over the variational ground state
${|G\rangle=\prod_i(\alpha+i\sqrt{1-\alpha^2}d^\dagger_{iy})|0\rangle}$,
see SM~\cite{SM} for details, leads to a free boson model
\begin{equation}
\label{eq:sw}
\tilde{\mathcal{H}}_\nu=\mu_\nu\! \sum_{i} 
\tilde{n}_{i\nu}
-\!
\sum_{ij}
\Big[
\mct_\nu\td_{i\nu}^{\dagger}\td_{j\nu}^{\phantom\dagger}-
\frac{\mcw_\nu}{2}
(
\td_{i\nu}^{\dagger}\td_{j\nu}^{\dagger}+ 
H.c.
)
\Big].
\end{equation}
Note that the parameters of this effective model in the ordered phase, given in Table~\ref{tab:sw}, depend on the flavor $\nu=x,y$.
The elementary excitations of (\ref{eq:sw}) have the dispersion
$\epsilon_\nu(\bk)=\sqrt{(\mu_\nu-2\mct_\nu\delta(\bk))^2-(2\mcw_\nu\delta(\bk))^2}$
with $\delta(\bk)=\cos k_x+\cos k_y$. 
In the $U^2(1)$ case with $\mcw=0$ both $x$ and $y$
modes are gapless with sound velocities 
$v_\nu\equiv\nabla_\bk\epsilon_\nu(\bk=0)
=\sqrt{8|\mcw_\nu|(\mct_\nu+|\mcw_\nu|)}$
vanishing at the transition.
Finite $\mcw$ in the $U(1)$ case leads to opening of a gap for $y$-excitations. The ratio
of the spectral weights of $I-$ and $R-$ propagators corresponding to $\chi^{II}_{yy}$ and
$\chi^{RR}_{yy}$ at $\Gamma$ point is given by~\cite{SM}
\begin{equation*}
\frac{\operatorname{Im}\chi^{II}_{yy}(0,\nu_\text{gap})}{\operatorname{Im}\chi^{RR}_{yy}(0,\nu_\text{gap})}\approx \frac{4\mcw}{(2\mct+\mcv)\phi^2},
\end{equation*}
which supports the interpretation that a dominant Hamiltonian term ($\mcw$) favors the amplitude fluctuations, while a dominant Weiss field ($\sim\mct\phi$) favors the gapped Goldstone fluctuations.

Finally, we address the behavior of the spin susceptibility $\chi^{SS}_{zz}$ in Fig.~\ref{fig:spin}
We observe that replacing the operator $d_{iy}$ 
in the strong-coupling expression for $S^z_i$ by its finite PEC value 
yields $S^z_i\sim(d_{ix}^{\dagger} +d_{ix}^{\phantom\dagger})\phi/2$.
In the ordered phase, the spin susceptibility $\chi^{SS}_{zz}$ therefore follows $\chi^{RR}_{xx}$, while they are decoupled in the normal phase. 

  

\begin{figure}[t]
\centering
\includegraphics[width=0.5\textwidth]{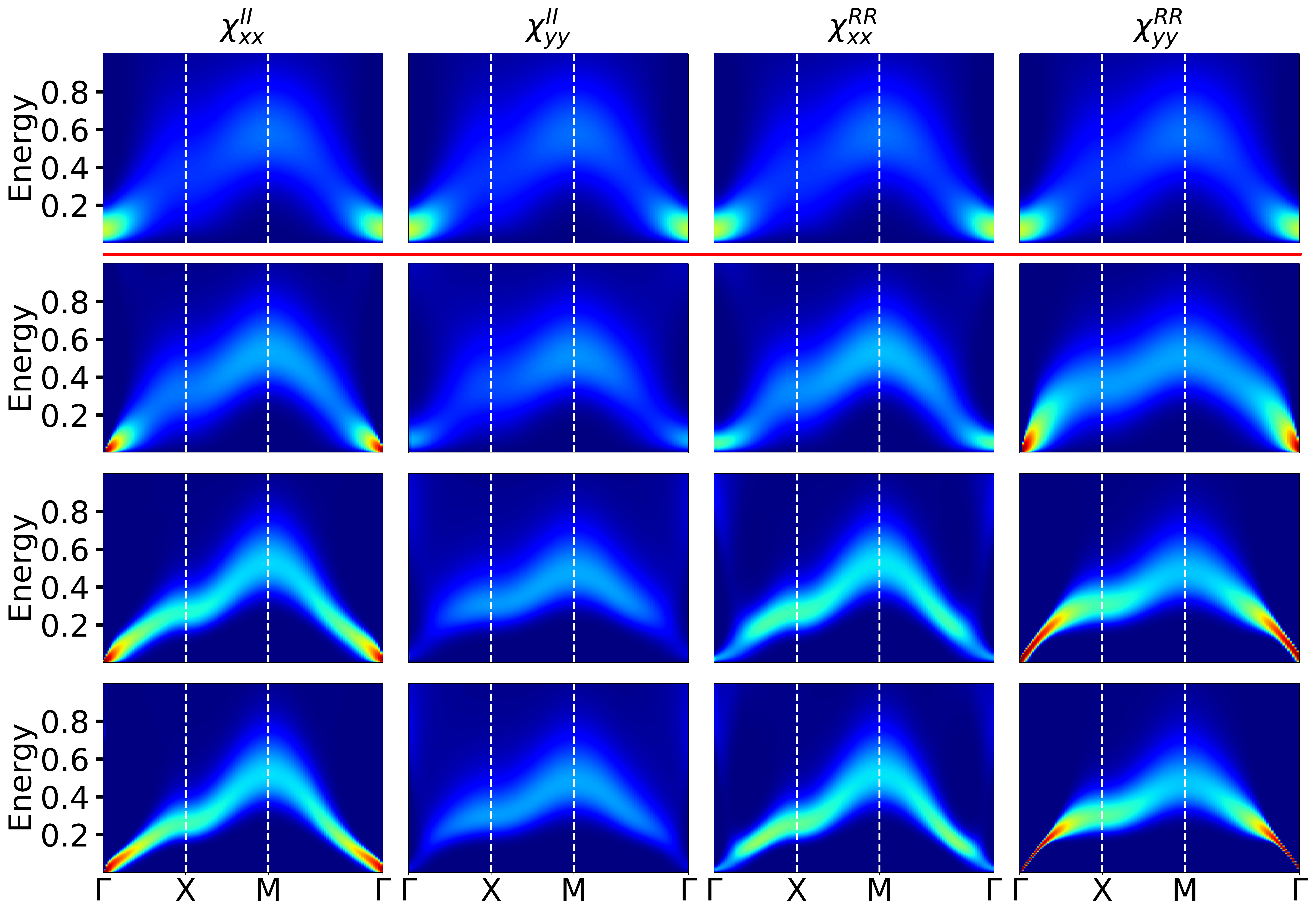}
\caption{\label{fig:temp} The same susceptibilities of $U^2(1)$ model as in Fig.~\ref{fig:goldstone} 
calculated across the thermally driven transition for $\Delta=$3.55.
The rows from top to bottom correspond to temperatures 
$T=1 / 11$, $1 / 16$, $1 / 30$, $1/40$ with $T_c\approx 1 / 13$.}
\end{figure}

\begin{figure}[t]
\centering
\includegraphics[width=\columnwidth]{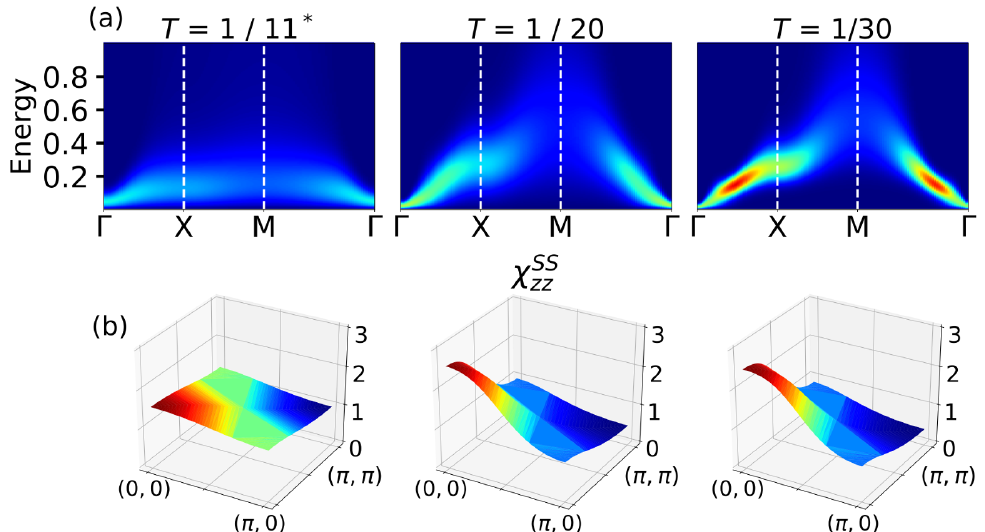}
\caption{\label{fig:spin_temp} (a) Evolution of dynamical spin susceptibility
$-\operatorname{Im}\chi^{SS}_{zz}(\bk,\omega)$ across the thermally driven phase
transition as in Fig.~\ref{fig:temp} for temperatures
$T=1/11$, $1/20$, $1/30$ (Asterisk marks the normal phase). 
(b) The corresponding static
susceptibilities $\operatorname{Re}\chi^{SS}_{zz}(\bk,0)$
throughout the Brillouin zone.}
\end{figure}

{\it Thermally driven transition.}
Since the transition observed in Pr$_{0.5}$Ca$_{0.5}$CoO$_3$~\cite{tsubouchi2002} is driven by temperature
we investigate the behavior of the  $U^2(1)$ model along the vertical trajectory in
Fig.~\ref{fig:1p_disp}b. We observe that the 1P gap in the normal state is closed, Fig.~\ref{fig:1p_disp}d.
The excitonic susceptibilities possess a peak at finite frequency, whose tail extends to zero frequency, Fig.~\ref{fig:temp}
Cooling is accompanied by a downward shift of the damped dispresive features, i.e.,
the phase transition can be viewed as a mode softening, an observation also made experimentally on TlCuCl$_3$~\cite{Merchant2014}.

The normal state spin susceptibility $\chi^{SS}_{zz}$ in Fig.~\ref{fig:spin_temp} does not
vanish as in Fig.~\ref{fig:spin}.
The presence of thermally excited HS states gives rise to $\bk$-featureless susceptibility with spectral
weight concentrated at low energies. Nevertheless, $\chi^{SS}_{zz}(\bk,\omega)$
changes qualitatively at the transition in this case as well. The
dispersion becomes sharper and its bandwidth increases significantly. As a result, upon cooling below $T_c$, the low-energy region is depleted of spectral weight throughout the Brillouin zone,
except in the vicinity of the $\Gamma$-point. 
Recently, this behavior was reported in inelastic neutron scattering
in the putative excitonic material
(Pr$_{1−y}$Y$_y$)$_{1−x}$Ca$_x$CoO$_3$~\cite{Moyoshi2018}.


In conclusion, we used DMFT to study the 2P response 
across exciton condensation transition in two-orbital Hubbard model. 
We observed the formation of GMs as predicted by symmetry considerations~\cite{Watanabe2012}.
Explicit breaking of continuous symmetry led to appearance 
of a gapped mode~\cite{Pekker2015}, characterized by vanishing 
of the gap at the phase transitions similar to observations in TlCuCl$_3$~\cite{Merchant2014}.
We have observed that the character of this mode changes from
Higgs-like amplitude fluctuations close to the phase boundary, to Goldstone-like
phase fluctuations deep in the ordered phase. We suggest that this behavior shall
be common to systems with weakly broken symmetry and provide an interpretation in terms of 
the relative strengths of the spontaneously generated Weiss field and the explicit
symmetry-breaking term in the Hamiltonian. 

Experimental observation of excitonic modes is in principle possible~\cite{Wang2018,Kim2014} using
resonant inelastic x-ray scattering, however, practical limitations in energy resolution 
and $\bk$-space accessibility~\cite{Wang2018} exist at the moment.
We have shown that the measurement of dynamical spin susceptibility provides an alternative,
that can used to identify spinful excitonic condensates with current experimental technology.

\begin{acknowledgements}
We thank J. Chaloupka, G. Sangiovanni, G. Khaliullin and B. Hartl for
discussions, H. Shinaoka for help with the ALPS code, K. Steiner and O. Janson for testing
the maximum entropy code, and K. Held and
A. Kauch for critical reading of the manuscript.
This work was supported by the ERC Grant Agreements No. 646807 under
EU Horizon 2020 (J.Ku., A.Har. and D.G.) and No. 306447 under EU
Seventh Framework Program
(FP7/20072013)/ERC (J.Ka.), by the Czech Ministry of Education, Youth
and Sports, project 
``IT4Innovations National Supercomputing Center – LM2015070'' (D.G.), 
by FWF through SFB ViCoM F41 (J.Ka.), by DFG through SFB1170
“Tocotronics” (A.Hau.) and by the Austrian Federal Ministry of Science, Research  and  Economy 
through the Vienna Scientific Cluster (VSC) Research Center (P.G.). We
gratefully acknowledge the Gauss Centre for Supercomputing
e.V. (www.gauss-centre.eu) for providing computing time on the GCS
Supercomputer SuperMUC at Leibniz Supercomputing Centre (www.lrz.de),
the programme ``Projects of Large Research, Development, and
Innovations Infrastructures'' (CESNET LM2015042) for  the access to
computing and storage facilities of the Czech National Grid
Infrastructure MetaCentrum, and the Vienna Scientific Cluster (VSC)
for access to its computing facilities.
\end{acknowledgements}

%
\end{document}